# Formation of convective cells in the scrape-off layer of the CASTOR tokamak

J Stöckel[1], P Devynck[2], J Gunn[2], E Martines[3], G Bonhomme[4], G Van Oost[5], M Hron[1], J Adamek[1], J Brotánková[1], R Dejarnac[1], I Duran[1], T Görler[6], T Hansen[5], R Panek[1], P Stejskal[1], V Svoboda[7], F Zacek[1]

[1] Institute of Plasma Physics, Association Euratom/IPP.CR, AS CR, Prague, Czech Republic
[2] Association Euratom/CEA sur la fusion controlèe, Saint Paul Lez Durance, France
[3] Consorzio RFX, Associazione Euratom/ENEA sulla Fusione, Padova, Italy
[4] LPMI, UMR 7040 du CNRS, Université Henri Poincaré, Vandoeuvre-lès-Nancy, France
[5] Department of Applied Physics, Gent University, Gent, Belgium
[6] University of Ulm, Germany
[7] Faculty of Nuclear Science and Physical Engineering, Czech Technical University, Prague

**Abstract**

We describe experiments with a biased electrode inserted into the scrape-off layer (SOL) of the CASTOR tokamak. The resulting radial and poloidal electric field and plasma density modification are measured by means of Langmuir probe arrays with high temporal and spatial resolutions. Poloidally and radially localized stationary structures of the electric field (convective cells) are identified and a related significant modification of the particle transport in the SOL is observed.

**1. Introduction**

Understanding and control of the particle and heat transport at the tokamak edge is of primary importance for design of future large-scale experiments. However, requirements for the transport properties of the plasma inside and outside the last closed magnetic surface are contradicting. A low heat conductivity and consequent steep gradients of plasma pressure must be kept inside the LCFS to achieve the best confinement. On the other hand, flat radial profiles (i.e. an amplified perpendicular transport) are required outside the LCFS (in the SOL) to reduce the power density deposited to the first wall elements (limiters, divertor plates...) thereby increasing their lifetime.

It is recognized that the edge transport inside the Last Closed Flux Surface (LCFS) can be efficiently controlled by imposing electric fields, as reviewed for tokamaks in [1,2] and for stellarators in [3]. In this case, an electrode is inserted inside the LCFS and biased with respect to the tokamak vessel. The magnetic surface, associated with the electrode position is biased and a strongly sheared radial electric field $E_r$ is formed between the electrode and the LCFS. Consequently, the plasma slab between the electrode and LCFS is forced to rotate poloidally due to the $E_r \times B_t$ drift. When the rotation is sheared, the plasma turbulence, which is the dominant mechanism behind the anomalous transport in this region, is reduced. The edge transport barrier, characterized by a steep gradient of plasma density is formed in front of the LCFS.

Experiments with manipulation of fluctuations (and transport) inside the SOL are scarcer. A signature of broadening of the SOL profiles has been achieved by biasing a divertor plate in the JFT-2M tokamak [4]. Toroidally elongated structures of plasma potential have been formed in the SOL and the associated $E_\perp \times B_t$ drift has been suggested as a mechanism of an enhanced transport [5]. Similar experiments were recently performed on the MAST spherical tokamak [6], where the divertor plates were biased as well. The formation of convective cells and a broadening of the power density profiles in the SOL were experimentally demonstrated [7].

Here, we present results of experiments at biasing of the electrode immersed into the SOL of the CASTOR tokamak. The diagnostics and magnetic configuration of the SOL is described in Sec. 2. Formation of convective cells is demonstrated in Sec. 3.

## 2 Arrangement of the experiment

**2.1 Diagnostic tools:** The experiments have been carried out on the CASTOR tokamak, which is a small size torus with a major radius of 0.4 m and a minor radius of 0.1 m. The toroidal magnetic field is 1.3 T and the plasma current is varied between 5 - 10 kA. The poloidal limiter with the radius a = 58 mm is equipped with 124 Langmuir tips, uniformly distributed around its circumference (see Fig. 1). The individual tips are spaced poloidally by ~3 mm.

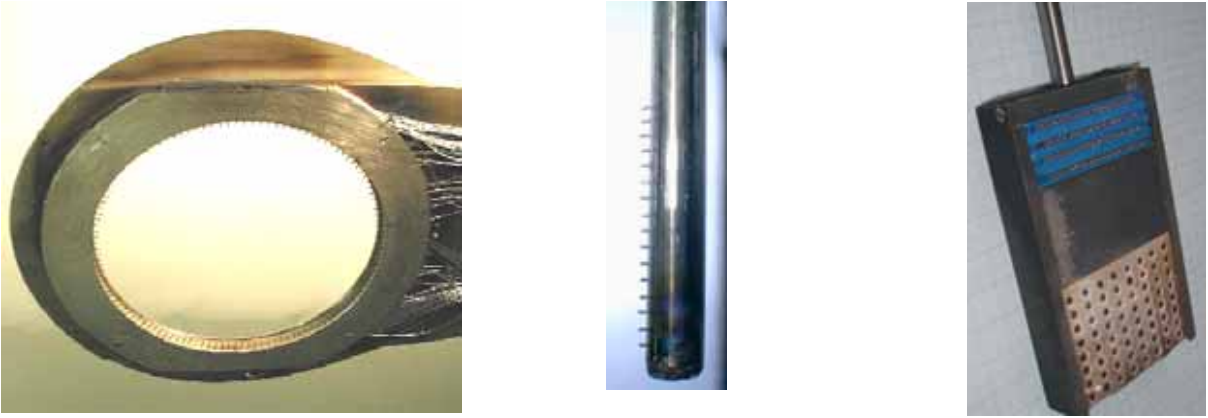

**Fig 1.** *Langmuir probe arrays for spatially resolved measurements. Left - the poloidal limiter equipped with 124 probes; Middle - the rake probe, Right – the 2D matrix of 64 probes.*

The rake probe is used to measure radial profiles of the edge plasma parameters in a single shot. The probe head is composed of 16 tips spaced radially by 2.5 mm and inserted into the edge plasma from the top of the torus $180^o$ toroidally away from the poloidal limiter. The two dimensional matrix of Langmuir probes can be inserted into the edge plasma to measure profiles in poloidal and radial direction simultaneously.

The tips of probe arrays measure either the floating potential or the ion saturation current. The floating potential $U_{fl}$ is related to the plasma potential $\varphi$ by the formula $U_{fl} = \varphi - \alpha T_e$. Assuming a flat electron temperature profile, which is reasonable for the edge region of the CASTOR tokamak, gradients of floating potential can be considered as indicative of electric fields. Similarly, the ion saturation current gradient gives an estimate of the local plasma density gradient.

2.2. **Asymmetry of the edge plasma:** Simultaneous measurements with the radial and poloidal probe arrays allow estimate asymmetries of the plasma column within the vacuum vessel of CASTOR. The distribution of the time averaged floating potential in the radial direction measured by the rake probe during the ohmic phase of a discharge is shown in the left panel of Fig. 2.

As seen from the figure, the floating potential exhibits a maximum. The maximum floating potential marks the location where a ExB velocity shear is present. Therefore, it is assumed that such maximum is located in the proximity of the last closed flux surface (LCFS) [1]. The radial position of the LCFS at the top of the torus is at ~ 45 mm in this particular case, which is noticeably deeper than the leading edge of the poloidal limiter a= 58 mm. Assuming a circular shaped plasma cross section, the observed difference between the LCFS and limiter radius at the top of the torus is interpreted as a downward shift of the plasma column by about 6-7 mm. It should be noted that previous measurements show that the floating potential decreases further and becomes negative at deeper insertions of the probe [1].

The right panel of Fig. 2 shows the distribution of the floating potential along the poloidal ring. As seen, the distribution at the top of the torus is relatively uniform, which means that the probes are located roughly on the same magnetic surface. On the other hand, the potential of the probes located at the bottom quarter of the poloidal circumference (in the range of poloidal angles $200^o$ - $320^o$) is negative. The sign of the phase velocity of the turbulence computed from the cross-correlation of couples of probes adjacent to each other, also shown in Fig. 2, clearly shows that these latter probes are located on the opposite side of the velocity shear than the other ones. This suggests that they are effectively deeper than the LCFS.

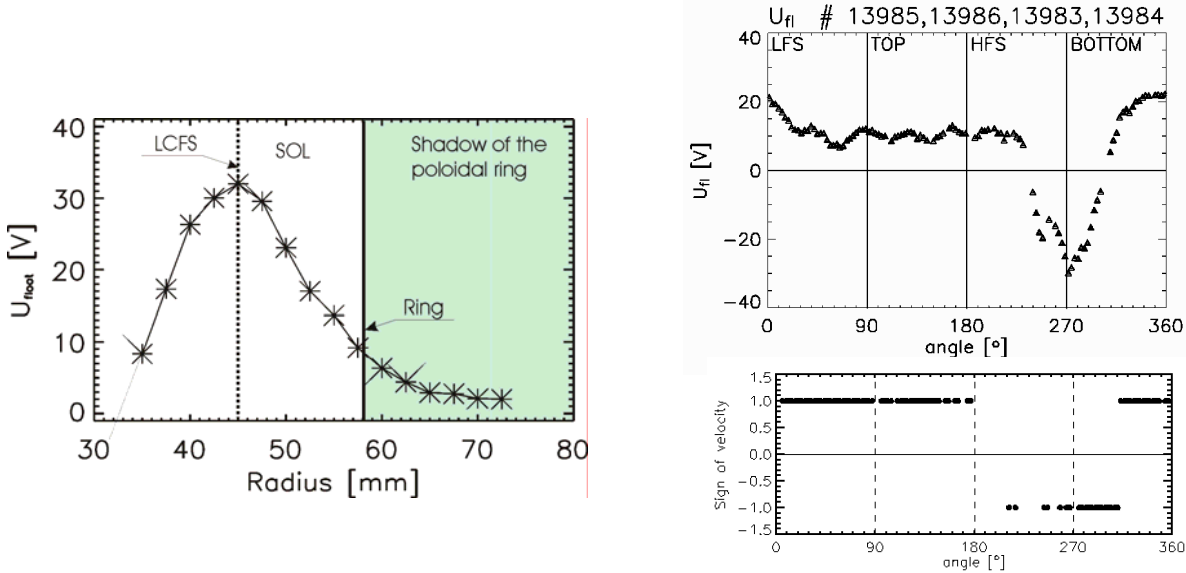

**Fig. 2.** *Left- Radial profile of the mean floating potential measured at the top of the torus. Right- Poloidal distribution of floating potential and of the sign of the turbulence phase velocity as measured by the poloidal ring in four reproducible shots. The positive sign of the $E_r \times B_t$ velocity corresponds to the poloidal propagation of turbulent structures in the region where $E_r>0$, i.e inside the SOL.*

These measurements are interpreted as follows. The SOL plasma is divided into two regions, regarding the parallel connection length *L* to a material surface, which is represented by the poloidal limiter:

- *Limiter shadow* is the region between the chamber wall and the leading edge of the poloidal limiter (58 mm < r < 100 mm). The corresponding connection length is about one toroidal circumference, L~ $2\pi R$.
- *Scrape-off layer* (SOL) with a much longer connection length is formed at the upper part of the plasma column because of the vertical shift of the plasma. The connection length L~ $q2\pi R$ is proportional to the local safety factor q (typically q = 6-9 in CASTOR). The radial extent of the SOL is largest at the top of the torus and depends on the value of plasma displacement.

**3. Biasing of the scrape-off layer**

The magnetic configuration described above is employed for the formation of stationary convective cells in the SOL of the CASTOR tokamak. The biasing electrode is immersed into the scrape-off layer from the top of the torus, 80° toroidally away from the limiter. The resulting positioning of the plasma column with respect to the poloidal limiter is schematically depicted in Fig. 3.

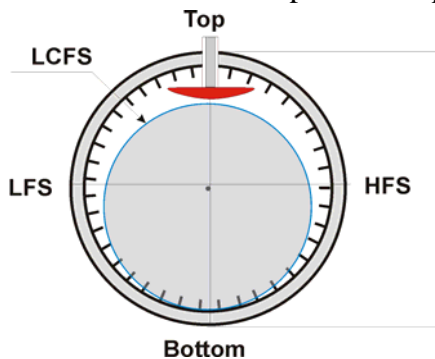

**Fig. 3** *Schematic picture of the poloidal cross-section of the CASTOR tokamak showing the respective position of the biasing electrode, plasma column and the poloidal limiter. The biasing electrode is located in the SOL.*

The graphite electrode has a mushroom-like shape. Its poloidal extent is 50 mm and the total collecting area ~ 15 cm$^2$. The electrode is biased positively with respect to the tokamak vessel. The typical current to the electrode is in the range of 20-40 A at the biasing voltage $V_b$ = 100-200 V.

A rather complex picture is observed when the electrode is biased. This is apparent from Fig. 4, where the poloidal distribution of the mean floating potential in ohmic and biasing phase of a discharge is compared for $V_b$= 150 V.

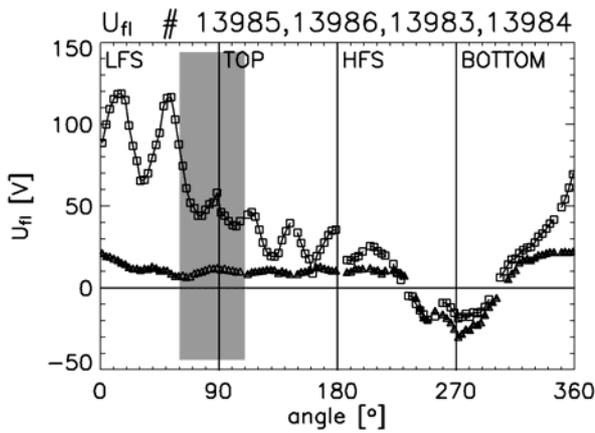

***Fig4.*** *Poloidal distribution of the mean floating potential $U_{fl}$ along the ring in ohmic (triangles) and biasing (squares) phase, as measured in four reproducible discharges with a constant edge safety factor $q(a)$~8. The poloidal position of the biasing electrode is marked by the bar.*

It is seen that the whole upper part of the torus is biased with respect to the ohmic level. Moreover, a strong poloidal modulation of the floating potential is observed in the range of poloidal angles $\theta = 0^o$-$200^o$. The observed peaks are interpreted as a signature of a biased flux tube, which originates at the electrode and extends along the helical magnetic field line. The electrode current flows predominantly parallel to the magnetic field lines in the upstream and downstream direction and terminates on the electron and ion side of the bottom part of the poloidal limiter, as it is apparent from the schematic picture, shown in the left panel of Fig. 5, which shows an unfolded magnetic surface associated with the biasing electrode. The projection of the biased tube on a poloidal plane is schematically depicted in the right panel of Fig. 5.

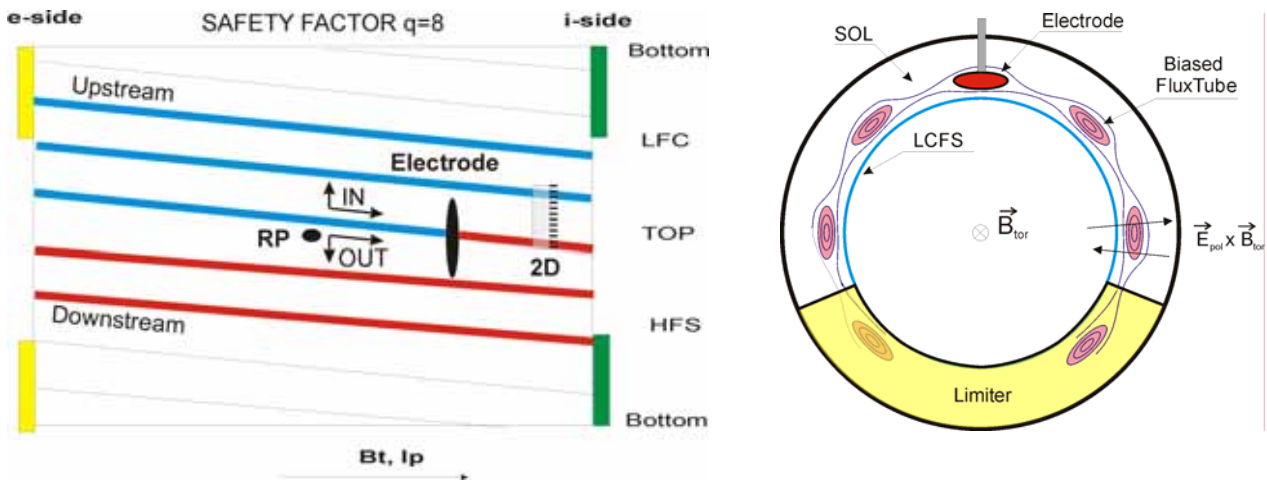

***Fig5.*** *Biased flux tube originating at the electrode is shown schematically for $q_{edge}$ ~ 8. Left – unfolded magnetic surface in the SOL. Location of the rake probe and of the 2D matrix is marked. Right - Projection of the tube on a poloidal cross section. Some equipotential surfaces are shown.*

The two dimensional character of the biased flux tube is directly observed by the 2D matrix of Langmuir probes inserted to the edge plasma from the top, $40^o$ away from the biasing electrode, as shown schematically in Fig. 5 – left. Its active surface (equipped with probes) is oriented downstream to the toroidal magnetic field lines. Projection of the biased flux tube on the 2D matrix is shown in Fig. 6.

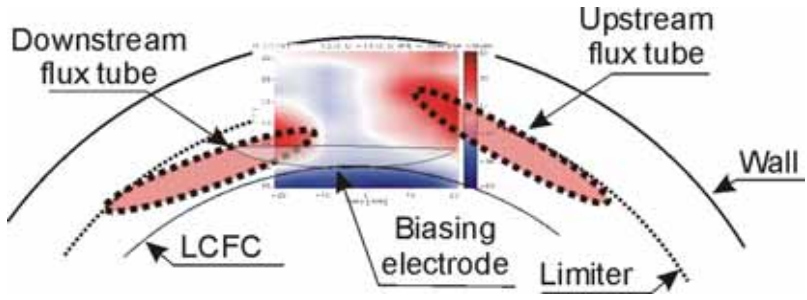

*Fig6. Poloidal projection of the biased flux tube as measured by the 2D matrix of Langmuir probes. The lengths of the downstream and upstream tubes are ~30 cm and 220 cm, respectively.*

The resulting perpendicular electric field is two-dimensional, having both radial and poloidal component. The amplitude and even the sign of the poloidal electric field $E_{pol}$ changes with the poloidal angle. A possible consequence of this is the occurrence of a convective motion of the SOL plasma because of the $E_{pol} \times B_t$ drift directed either inward or outward according the sign of the $E_{pol}$. In order to detect this effect, the tips of the poloidal ring have been used to measure either the floating potential or the ion saturation current in two subsequent discharges. A detailed distribution of these two quantities along one quarter of the poloidal ring is shown in Fig. 7.

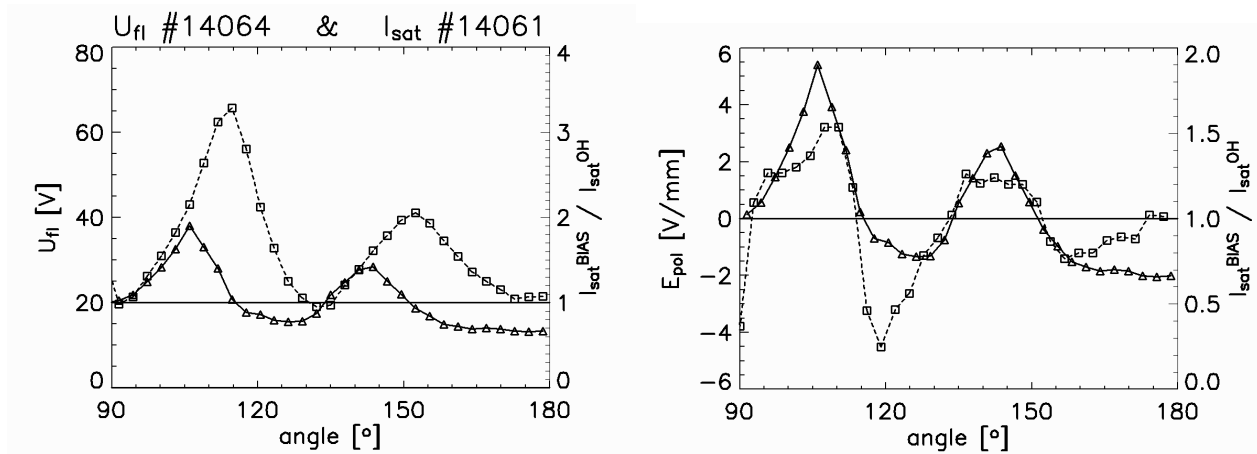

**Fig. 7.** Left - *Poloidal distribution of the floating potential (squares) and of the ion saturation current (triangles) along one quarter of the poloidal ring (High Field Side – top). Right - Distribution of the poloidal electric field (squares) and the ion saturation current (triangles) along one quarter of the poloidal ring. The ion saturation current is normalized to its ohmic value.*

It is seen from the figure that the poloidal distribution of the ion saturation current (proportional to the plasma density) is modulated by the biasing with the same periodicity as the potential distribution, but is shifted poloidally. The ion saturation current (plasma density) increases at one side of the potential hills, while it is reduced below the ohmic level on the other side. Furthermore, the poloidal electric field has been calculated as the difference of the floating potentials of adjacent tips divided by their distance and compared with the poloidal distribution of the plasma density in the right panel of Fig. 7. It is evident that the relative change of the plasma density is proportional to the local value of the poloidal electric field. Consequently, the radial particle flux driven by the poloidal electric field, easily computed as

$$\Gamma \cong nv_{ExB} = \frac{E_{pol} n}{B_t} \qquad (1)$$

is directed either outward or inward, depending on the sign of the poloidal electric field. Its peak value (1022 m-2s-1) is almost by two orders of magnitude higher than the poloidally averaged particle flux through the LCFS, estimated from the global particle balance. This implies that the biasing is indeed having a very relevant effect on the overall particle balance of the SOL.

The strong effect of biasing on the SOL equilibrium profile is further confirmed by measurements of the radial profile of the ion saturation current, obtained using the rake probe. Such measurements are shown in Fig. 8, both for the ohmic phase and during the SOL biasing.

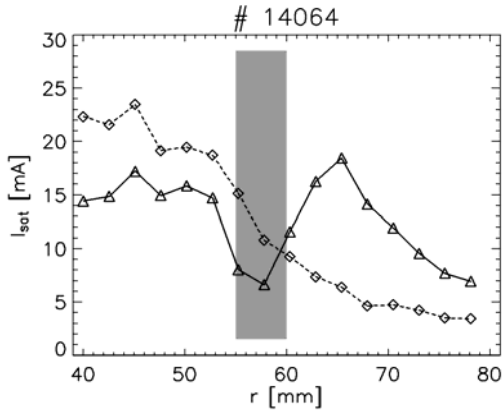

**Fig. 8** *Comparison of the radial profile of the ion saturation current in the ohmic (dashed line) and in the SOL biasing (solid line) phase of the discharge. The gray bar marks the electrode position. The rake probe is located poloidally at θ ~ 90º(see Fig. 5), which corresponds to a region of positive electric field (outward particle flux) during the biasing phase.*

The density profile displays a strong depletion at the location of the electrode during the biasing phase. The density in front of this position is also reduced, whereas a density accumulation can be seen in the region between 60 and 70 mm, behind the electrode. This strong modification of the density profile is consistent with an increase of the particle flux at the radial location where the electrode is placed.

In fact, the density profile can be expected to be different at different poloidal locations, due to the poloidal modulation of the particle flux. However, according to the tracing of the magnetic field lines the rake probe is located in a region of positive electric field, and therefore of positive particle flux, so that the effect observed on the density profile is indeed consistent with the local value of the particle flux.

**4 Conclusions**

With the help of a full poloidal ring of Langmuir probes it has been possible to demonstrate the formation of stationary convective cells during SOL biasing in CASTOR. These are formed by the creation of a biased flux tube, which makes several toroidal turns in the SOL, according to the local safety factor value. As a consequence, not only the radial, but also a rather strong poloidal electric field is formed at the magnetic surface associated with the biased electrode. This poloidal field changes its sign periodically along the poloidal circumference. Furthermore, the density is also poloidally modulated. The overall result is the creation of a pattern of particle flux with a strong modulation that makes it negative at some poloidal locations. This modulation can be seen as the formation of convective cells around the biased flux tube.

The results described in this paper are related to the SOL geometry of the CASTOR tokamak. In the present configuration, the downward shift of the plasma column allows flux tubes in the top part of the machine to describe several turns around the torus, which is an essential feature for the formation of the convective cells. Nevertheless, it is easy to envisage the application of the same technique to other geometries, including that of divertors.

It is important to emphasize that the technique described herein allows a strong increase of the transport in some regions of the SOL by the use of a single electrode having a limited extension in space. These results could have practical consequences for SOL engineering and hence for exhaust in large scale devices.

**Acknowledgement**: Authors are indebted to F Jiranek, V Havlik, K Rieger, M Satava and J Zelenka for the design, construction of diagnostics and technical assistance in the experiment. This work has been carried out with the support of the projects 202/03/0786 and 202/03/P062 (Grant Agency of the Czech republic) and No. 2001-2056 (INTAS).

**References**
[1] Van Oost G, Stockel J, Hron M, Devynck P, Dyabilin K, Gunn J, Horacek J, Martines E, Tendler M 2001 J. Fusion Phys Res. SERIES **4** 29
[2] Van Oost G et al 2003 Plasma Phys. Control. Fusion **45** 621
[3] Hidalgo C et al in Proc. of 30th EPS Conf on Contr. Fusion Plasma Phys, St Petersburg, July 2003, ECA Vol 27A, P-1.21
[4] Hara J et al 1997 Journal of Nuclear Materials **241-243** 338
[5] Krlin L, Stockel J, Svoboda V 1999 Plasma Phys. Control. Fusion, **41** 339
[6] Ryutov D D, Helander P, Cohen RH 2001 Plasma Phys. Control. Fusion, **43** 1399
[7] Counsell GF, Alm JW, Cohen RH, Kirk A, Helander P, Martin R, Ryutov DD, Tabasso A, Wilson HR, Yang Y Nuclear Fusion 2003, **43** (10): 1197-1203